\documentclass {pasj00}

\begin{document}
\SetRunningHead{Nagai et al.}{VLBI Monitoring of 3C~84 in Early Phase of the 2005 Outburst}

\title{VLBI Monitoring of 3C~84 (NGC~1275) in Early Phase of the 2005 Outburst}


\author{Hiroshi \textsc{Nagai},\altaffilmark{1} \thanks{hiroshi.nagai@nao.ac.jp}
        Kenta \textsc{Suzuki},\altaffilmark{2, 1}
	Keiichi \textsc{Asada},\altaffilmark{3}
	Motoki \textsc{Kino},\altaffilmark{1}
	Seiji \textsc{Kameno},\altaffilmark{4}
	Akihiro \textsc{Doi},\altaffilmark{5}
	Makoto \textsc{Inoue},\altaffilmark{3}
	Jun \textsc{Kataoka},\altaffilmark{6}
	Uwe \textsc{Bach},\altaffilmark{7}
	Tomoya \textsc{Hirota},\altaffilmark{1}
	Naoko \textsc{Matsumoto},\altaffilmark{8, 1}
	Mareki \textsc{Honma},\altaffilmark{1}
        Hideyuki \textsc{Kobayashi},\altaffilmark{1}
	and
	Kenta \textsc{Fujisawa}\altaffilmark{9}
        }
\altaffiltext{1}{National Astronomical Observatory of Japan, Osawa 2-21-1, Mitaka, Tokyo 181-8588}
\altaffiltext{2}{Department of Astronomy, Graduate School of Science, The University of Tokyo, 7-3-1 Hongo, Bunkyo-ku, Tokyo 113-0033}
\altaffiltext{3}{Institute of Astronomy and Astrophysics, Academia Sinica. P.O. Box 23-141, Taipei 10617, Taiwan, R.O.C.}
\altaffiltext{4}{Faculty of Science, Kagoshima University, 1-21-35 Korimoto, Kagoshima 890-0065}
\altaffiltext{5}{Institute of Space and Astronautics, Japan Aerospace Exploration Agency, 3-1-1 Yoshinodai, Sagamihara, Kanagawa 229-8510}
\altaffiltext{6}{Research Institute for Science and Engineering, Waseda University, 3-4-1, Okubo, Shinjuku, Tokyo, 169-8555}
\altaffiltext{7}{Max-Planck-Institut f\"{u}r Radioastronomie, Auf dem H\"{u}gel 69, 53121 Bonn, Germany}
\altaffiltext{8}{Department of Astronomical Science, The Graduate University for Advanced Studies (SOKENDAI), Osawa 2-21-1, Mitaka, Tokyo 181-8588}
\altaffiltext{9}{Faculty of Science, Yamaguchi University, 1677-1 Yoshida, Yamaguchi, Yamaguchi 753-8512}


%


\maketitle

\begin{abstract}
Multi-epoch Very Long Baseline Interferometry (VLBI) study of the sub-pc scale jet of 3C~84 is presented.  We carried out 14-epoch VLBI observations during 2006-2009 with the Japanese VLBI Network (JVN) and the VLBI Exploration of Radio Astrometry (VERA), immediately following the radio outburst that began in 2005.  We confirmed that the outburst was associated with the central $\sim1$~pc core, accompanying the emergence of a new component.  This is striking evidence of the recurrence of jet activity.  The new component became brighter during 2008, in contrast to the constant $\gamma$-ray emission that was observed with the $Fermi$ $Gamma$-$ray$ $Space$ $Telescope$ during the same time.  We found that the projected speed of the new component is $0.23c$ from 2007/297 (2007 October 24) to 2009/114 (2009 April 24).  The direction of movement of this component differs from that of the pre-existing component by $\sim40^{\circ}$.  This is the first measurement of kinematics of a sub-pc jet in a $\gamma$-ray active phase.  Possible detection of jet deceleration and the jet kinematics in connection with the $\gamma$-ray emission is discussed. 
\end{abstract}

\section{Introduction}
The idea of the recurrence of jet activity in active galactic nuclei (AGN) has been proposed since early times (e.g., Burbidge \& Burbidge 1965; Kellermann 1966), but no one has identified the site of the recurrence directly.  The radio source 3C~84, associated with NGC~1275 ($z=0.0176$), shows multiple lobe-like features from pc to 10-kpc scale, suggestive of the recurrent jet activities (e.g., Pedlar et al. 1990).  In the pc-scale region, there is a pair of lobes which is probably formed by the jet activity originating in the 1959 outburst (e.g., Asada et al. 2006).  Recently, radio observations with a single dish telescope detected an outburst starting in 2005 (Abdo et al. 2009, hereafter A09).  The Very Long Baseline Array (VLBA) observations also found a radio brightening in the central sub-pc structure by comparing the data of two epochs between 2007 and 2008 (A09).  The radio core brightening is ascribed to a new episode of the recurrent activity.  However, it is still unclear whether the flare is indeed associated with the recurrence because of a lack of intensive Very Long Baseline Interferometry (VLBI) monitoring.  In late 2008, three years from the beginning of the outburst, the Large Area Telescope (LAT) of the $Fermi$ $Gamma$-$ray$ $Space$ $Telescope$ revealed GeV $\gamma$-ray emission from NGC~1275 (A09), which is about 7 times larger than the upper limit constrained by the $EGRET$.  This radio core brightening coinciding with the $Fermi$ $\gamma$-ray detection may indicate that the $\gamma$-ray emission is associated with the core of 3C~84.  What happened at the early stage of the outburst is of great interest.

In the present paper we report the results from a combination of the archival data and new monitoring observations of 3C~84 in progress.  Our data allows us to resolve $\sim1$~pc core during the epochs closer to the trigger of radio flare than those of A09.  
Throughout this paper, we adopt $H_{0}=71$~km sec$^{-1}$ Mpc$^{-1}$, $\Omega_{\mathrm{M}}=0.27$, and $\Omega_{\mathrm{\Lambda}}=0.73$ (1~mas=0.353~pc, 0.1~mas~yr$^{-1}=0.113c$). 

\section{Observation and Data Reduction}
\subsection{Archival Data}
We used archival data of VLBI Exploration of Radio Astrometry (VERA) from 2006 May to 2008 May (2006/134, 2006/143, 2006/346, 2007/142, 2007/258, 2007/297, 2007/324, 2007/361, 2008/035, 2008/063, 2008/106, 2008/141\footnote{Observing sessions are denoted by year/day of the year.}).  Observations were carried out with four VERA stations at 22.2~GHz, where 3C~84 was being used as a bandpass calibrator or a fringe finder.  Typically five scans of 5-minutes duration were obtained.  Left hand circular polarization (LHCP) was received and sampled with 2-bit quantization, and filtered using the VERA digital filter unit (Iguchi et al. 2005).  The data were recorded at a rate of 1024~Mbps, providing a bandwidth of 256~MHz in which 14 IF-channels per a total of 16 IF-channels of 16~MHz bandwidth were assigned to 3C~84.  For 2006/134, 2006/143, and 2007/297 data, only 1 IF-channel with 8~MHz bandwidth was assigned, and for 2007/142 data, 2 IF-channels with 8~MHz bandwidth were assigned.  Correlation processes were performed with the Mitaka FX correlator (Chikada et al. 1991).  

\subsection{New Observations with VERA and JVN}
The Japanese VLBI Network (JVN: Fujisawa 2008) observations were carried out on 2008/354 using the Tomakomai 11-m telescope and the four VERA stations at 22.2~GHz, and on 2008/356 using the Yamaguchi 32-m telescope, the Tsukuba 32-m telescope, and the VERA 4 stations at 8.4~GHz.  The VERA observation was carried out on 2009/114 at 22.2~GHz.  The Nobeyama 45-m telescope and the Kashima 34-m telescope also participated in the VERA observation.  Right hand circular polarization was received at 8.4~GHz, and LHCP was received at 22.2~GHz.  The data were recorded at the rate of 128~Mbps.  Data correlation was performed with the Mitaka FX correlator.  VERA observations were performed in dual-beam phase referencing mode, but in the present paper we report the analysis using the one-beam data.  The dual-beam analysis will be presented in an upcoming paper.

\subsection{Single-Dish Monitoring with Effelsberg}	
The flux density measurements at the Effelsberg 100-m telescope of the Max-Planck-Institut f\"{u}r Radioastronomie (MPIfR) were obtained during regular calibration and pointing observations at 22~GHz. The measurements were done using cross-scans in azimuth and elevation direction. The data reduction was done in the standard manner as described by Kraus et al. (2003).  The measured antenna temperatures were linked to the flux-density scale using primary calibrators like NGC~7027, 3C~286, and 3C~48 (Ott et al. 1994; Baars et al. 1977).

\subsection{VLBI Data Reduction}
Data reduction was performed using the NRAO Astronomical Imaging Processing System (AIPS).  A standard $a~priori$ amplitude calibration was performed using the AIPS task APCAL based on the measurements of the system temperature~($T_{\mathrm{sys}}$) during the observations and the aperture efficiency provided by each station, for VERA and JVN data at 22~GHz.  This flux calibration achieved accuracies of 10\% or less.  Such a calibration process was not used for JVN data at 8.4~GHz since some of the antennas were not equipped with the system of the $T_{\mathrm{sys}}$ measurement.  We adopted a calibration method using the flux calibrator described in Doi et al. (2007).  This flux calibration method can achieve accuracies of $\sim10$\% or less according to several JVN experiments.  A scaling factor of the amplitude was derived from comparison of the correlated flux of a compact radio source DA~193 and flux measurement by a single dish observation with the Yamaguchi telescope carried out 6 days after VLBI observation.  Since there is no time variation of the total flux density of DA~193 exceeding its error within a timescale of week by Mets{\"a}hovi observation at 22~GHz (Ter{\"a}sranta et al. 2005), we expect that the time variation during the epochs between the JVN observation and the Yamaguchi observation is not significant.  Total flux density of DA~193 was estimated to be $4.84$~Jy.  DA~193 is so compact that visibility amplitude of JVN baselines at 8.4 GHz can be calibrated using the total flux density.  DA~193 was observed at nearly the same elevation of 3C~84 for 10 minutes, and we applied the scaling factor to 3C~84 at all observing times assuming no significant time variation and elevation dependence of system noise equivalent flux density (SEFD) during the observation.  According to several JVN experiments, time variation of SEFD during the observation is typically 10\% at 8.4~GHz.  Therefore, we expect that the flux calibration accuracy is at most 20\%.  Fringe fitting was done using the AIPS task FRING.  Final images were obtained after a number of iterations with CLEAN and self-calibration using the Difmap software package (Shepherd 1997).


\section{Results}
\begin{figure*}
  \begin{center}
    \FigureFile(165mm,75mm){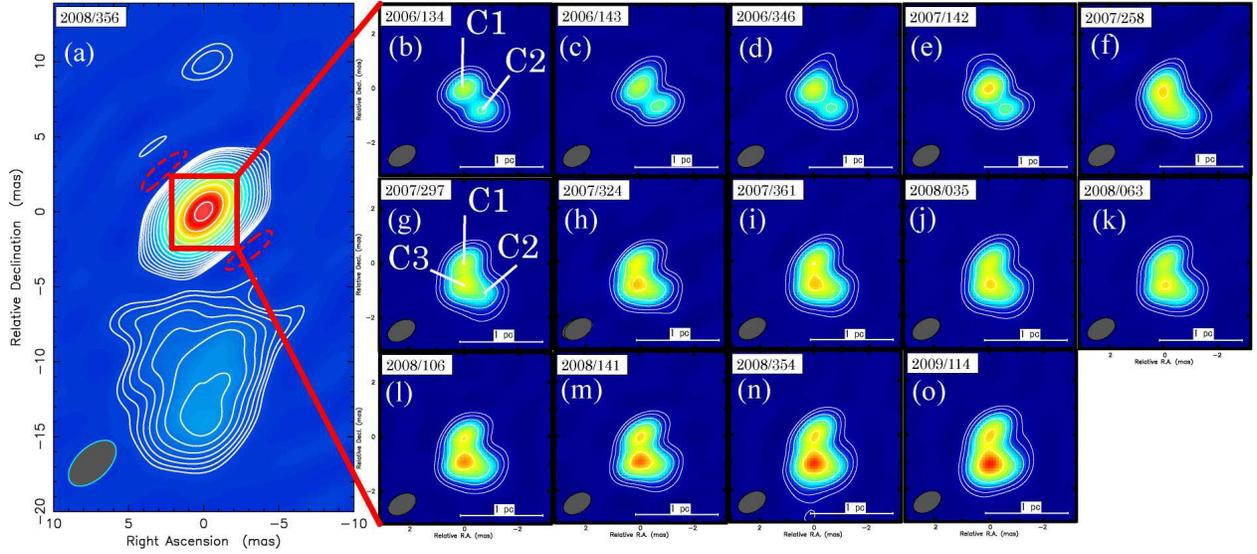}
  \end{center}
  \caption{(a) JVN image of 3C~84 at 8~GHz.  The contours are plotted at the level of $86.2$~mJy $\times\left( \sqrt{2} \right)^{n}$ ($n=-1, 0, 1, 2, 4, \dots, 128$).  {bf The lowest contour corresponds to three-times image noise r.m.s..}  The beam size is $3.85\times2.14$~mas at the position angle of $-47^{\circ}$, which is shown in the left corner of the image. (b)-(o) VERA images of 3C~84 at 22~GHz.  All images are shifted in reference to the northern component (component C1).  The contours are plotted at the level of 4.18, 8.36, 16.72, 33.44, and 66.88 \% of the peak intensity (4.989~Jy/beam) on 2009 April 24.  The restoring beam (1.1$\times$0.7~mas, position angle of $-60^{\circ}$) was set to make images uniform.}\label{fig:image}
\end{figure*}

Figure \ref{fig:image}(a) shows the pc-scale radio feature at 8.4~GHz.  A lobe-like feature to the south of the bright core is visible.  A faint counter-jet component is marginally detected.  Overall structures are similar to those by previous VLBA observation in $\gamma$-ray quiet phase (Walker et al. 2000).  For more on detailed structure at low frequencies, see Walker et al. (2000) and Asada et al. (2006).  Figure \ref{fig:image}(b)-(o) shows close-up images of the central $\sim1$~pc region at 22.2~GHz.  On 2008/354 Tomakomai was, and on 2009/114 Kashima and Nobeyama were involved in the observation in addition to VERA, but we produced these images by analysis with just VERA stations at this moment.  Full analysis including additional stations will be reported elsewhere.  

\begin{figure}
  \begin{center}
    \FigureFile(85mm,30mm){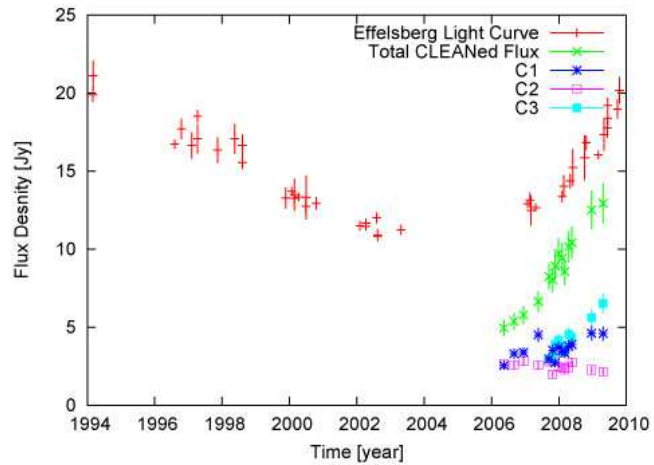}
  \end{center}
  \caption{Pluses: The Effelesberg light curve of 3C~84 at 22~GHz.  Crosses: total CLEANed flux of VERA observation at 22.2~GHz.  Asterisks:  The light curve of component C1.  Open squares: The light curve of component C2.  Filled squares: The light curve of component C3. }\label{fig:lightcurve}
\end{figure}

In Figure \ref{fig:lightcurve}, we show the light curve monitored with the Effelsberg 100-m telescope and total CLEANed flux at 22.2~GHz with VERA as a function of time.  It is obvious that the increase of total flux density correlates with that of total CLEANed flux, demonstrating that the outburst is associated with the central 1-pc region.       

In the first 4 epochs, component C2 was visible at the position separated by $\sim1$~mas from the central core (C1) in a position angle of $-141^{\circ}$.  The alignment of these components was similar to that in the $\gamma$-ray quiet phase (Dhawan et al. 1998).  During first 4 epochs, component C2 showed no significant motion relative to component C1.  Remarkably, a new component (C3) suddenly emerged to the south of the central core on 2007/258 (Fig. \ref{fig:image}(f)) despite only 4 months having passed since the previous observation (Fig. \ref{fig:image}(e)).  Moreover, component C3 was clearly resolved on 2007/297 (Fig. \ref{fig:image}(g)), separated by only one month from the previous epoch observation in which component C3 was blended with the other components.  In later epochs, component C3 appeared more significant.  The flux density of each component is also plotted in Figure \ref{fig:lightcurve}.  Component C3 showed the most significant increase in flux.
	

Figure \ref{fig:position} shows the separation between the components C1 and C3 as a function of time.  The position of the components C1 and C3 was derived from the two-dimensional Gaussian fit in the interferometric ($u, v$)-plane using the ``modelfit" task in Difmap.  It is difficult to measure the positional error of each component quantitatively from the interferometric data in each epoch independently.  We thus employed the method described in Homan et al. (2001).  We initially set the uncertainty for each data point equal to unity, and then we performed a linear fit to the data assuming the motion with constant speed to obtain a preliminary $\chi^{2}$.  Taking this preliminary $\chi^{2}$, we then uniformly rescaled the uncertainty of each data point such that reduced-$\chi^{2}$ to be unity.  Finally, the positional error of each data point is estimated to be 0.013~mas.  This error is typically two times larger than the one estimated from the signal-to-noise ratio (SNR) such that $\theta_{\mathrm{comp}}/\mathrm{SNR}$, where $\theta_{\mathrm{comp}}$ is the component radius.  This fit results in an apparent speed of 0.20$\pm$0.01~mas/yr (projected speed of $0.23\pm0.01c$) towards the south.  This is approximately consistent with the jet speeds in $\gamma$-ray quiet phase (Dhawan et al. 1998).    The direction of movement of the new component differs from the alignment of the components C1 and C2 by $\sim40^{\circ}$ on the projected plane.  We note that we did not include the data on 2007/258 to this fit because component C3 might have moved faster before 2007/297 (see \S 4).

\section{Discussion}
In the preceding section, we have shown the observational results of three components in the central $\sim1$~pc core.  In this section, we particularly focus on component C3.  Firstly, we will discuss the origin of component C3.  We consider that this component was ejected from the core as follows: (i) component C3 was ejected around 2007/142 (Fig. \ref{fig:image}(e)), but cannot be distinguished from the core (C1) because the component was in the close vicinity of the core, (ii) component C3 was distinguished from the core ($\sim0.7$~mas from the core; see Fig. 3) at epochs later than 2007/258 (Fig. \ref{fig:image}(f)) as a result of motion to the south.  This interpretation is supported by the following evidence.  The flux density of component C1 had increased until 2007/142 and then decreased on 2007/258 (see Fig. \ref{fig:lightcurve}), suggesting that component C3 was isolated from the core between 2007/142 and 2007/258.  Furthermore, component C3 indeed moved, separating from the core at epochs later than 2007/142.  Therefore, component C3 was likely to be ejected from the core triggered by a new episode of recurrent activity.  

\subsection{Deceleration or Absorption Effect?}
\begin{figure}
  \begin{center}
    \FigureFile(80mm,30mm){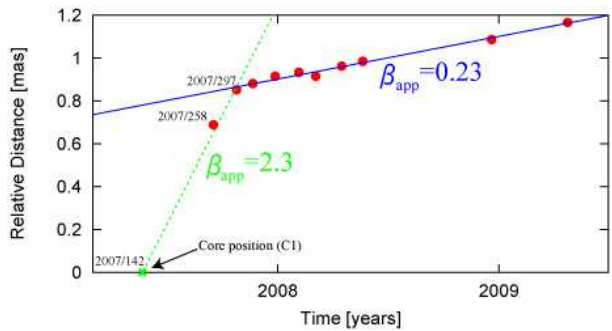}
  \end{center}
  \caption{Plot of the separation between component C3 and component C1.  The error bar is smaller than the size of each symbol.  The blue solid line represents a linear fit to the data from 2007/297 to 2009/114.  The green broken line represents the one from 2007/142 to 2007/297, assuming that component C3 was ejected from the position of component C1 on 2007/142 (see \S4.2).}\label{fig:position}
\end{figure}

Here we discuss interpretations of the apparent speed of component C3 (Fig. \ref{fig:position}).  If we assume that component C1 is stationary and component C3 was ejected from the position of component C1 on 2007/142, the apparent speed of component C3 is $\beta_{app}=2.3$ between 2007/142 and 2007/297 (green broken line in Figure \ref{fig:position}), yielding $\beta>0.92$ with the viewing angle of $<45^{\circ}$.  This implies that the jet starts with a relativistic speed and decelerates down to sub-relativistic speed over a projected distance of $\sim0.8$~mas ($\sim0.28$~pc).  A09 also modeled the observed SED with a decelerating jet model (Georganopoulos \& Kanzas 2003), which is developed by following the fact that VLBI studies found no superluminal motion in the jets of BL Lac objects (e.g., Edwards \& Piner 2002).  A fit using this model derives that the jet decelerates from $\Gamma=10$ ($\beta=0.995$) to $\Gamma=2.0$ ($\beta=0.87$) over a distance of 0.16~pc.  This is roughly consistent with our VLBI observation.  

In above discussion, we assumed that component C3 was ejected from the core (C1) on 2007/142, but we cannot completely exclude the possibility that component could be ejected before 2007/142.  Apparent absence of component C3 before 2007/297 could arise from the free-free absorption (FFA).  The evidence of the FFA in the pc-scale region of 3C~84 was firstly pointed out by Vermeulen, Readhead, \& Backer (1994) and Walker, Romney, \& Benson (1994).  Walker et al. (2000) revealed the distribution of the FFA by multi-frequency VLBA observation.  The absorption is greater around the radio core and falls off with distance.  The geometry is consistent with the absorption by ionized gas associated with the accretion disk.  In this model, the receding jet (northern jet) is obscured by the accretion disk.  However, the component in the close vicinity of the core can suffer from the FFA even in the approaching jet (southern jet) depending on the jet inclination angle and the thickness of the accretion disk (e.g., NGC~1052; Kameno et al. 2001).  Therefore, the apparent absence of component C3 in earlier epochs can result from the absorption via the accretion disk.  If this is the case, our estimated speed could be overestimated.  Furthermore, the synchrotron self-absorption is also possible to play a role for the reduction of flux density of component C3.  We are doing higher frequency observations to overcome these problems.  We will report the result in another paper.
	
\subsection{Break of One-zone Approximation}
As discussed in A09, the radio core brightening may be related to the GeV $\gamma$-ray emission detected with $Fermi$.  Our observation indicates that component C3 plays a main role in this radio brightening, allowing us to expect that this component is responsible for the $\gamma$-ray emission.  However, there is no obvious correlation between the radio light curve of component C3 and the $\gamma$-ray emission: component C3 showed radio brightening while the $\gamma$-ray showed no significant variation during 2008 (A09).  Furthermore, component C3 showed a sub-relativistic speed during 2008, while the one-zone synchrotron self-Compton (SSC) model derived a mildly relativistic flow ($\Gamma=1.8$, $\beta=0.83$) (A09).  Similar apparent slow-moving jet in spite of strong $\gamma$-ray emission is seen in M~87 (Kovalev et al. 2007; Ly et al. 2007) and Mrk~501 (Giroletti et al. 2004; Giroletti et al. 2008).   A spine sheath model has been suggested to ease this problem: a fast spine jet produces the strong TeV emission by inverse Compton scattering of the radio photon from a surrounding slow layer (Ghisellini et al. 2005).  Such a spine-sheath structure is clearly visible in these sources.
  It is difficult to confirm the spine-sheath structure in 3C~84 with our data because of lack of spatial resolution.  A higher angular resolution telescope, such as VSOP-2 (Tsuboi et al. 2009), will address this issue in the near future.

\section{Summary}
We have revealed that the recent 3C~84 radio outburst is associated with a new jet ejection.  The direction of movement of the newly ejected component differs from that of the pre-existing component by $\sim40^{\circ}$.  The new component shows a projected speed of $0.23c$ during the $Fermi$ observation.  The light curve of the new component shows no obvious correlation with the $Fermi$ $\gamma$-ray flux.  Although we need further careful study of the kinematics of the new component, this component might start with a possible relativistic speed ($\beta_{app}=2.3$) at first, and then decelerates down to a sub-relativistic speed.  Possible detection of the jet deceleration agrees with a decelerating jet model adopted to explain the $\gamma$-ray emission.  
		
\bigskip
We thank an anonymous referee for reading the manuscript and constructive comments.  We are grateful to the staff of all the VERA and JVN stations for their assistance in observations.  This work is based on observations with the 100-m telescope of the MPIfR at Effelsberg.  We thank Alex Kraus for providing us with additional archival Effelsberg data (1994-2004).  
This research has made use of the NASA/IPAC Extragalactic Database (NED) which is operated by the Jet Propulsion Laboratory, California Institute of Technology, under contract with the National Aeronautics and Space Administration. 


\end{document}